# Direct fabrication of high-quality ring-shaped REBa$_2$Cu$_3$O$_y$ bulk magnets by the Single-Direction Melt Growth (SDMG) method

T. Motoki,[1,a)] M. Semba,[1] and J. Shimoyama[1]

[1]*Department of Physical Sciences, Aoyama Gakuin University, Sagamihara, 252-5258, Japan*

[a)] Author to whom correspondence should be addressed.  Electronic mail:  motoki@phys.aoyama.ac.jp Tel.: 81 42 759 6185

Ring-shaped REBa$_2$Cu$_3$O$_y$ melt-textured bulks have been successfully grown by the Single-Direction Melt Growth (SDMG) method, which enables the direct fabrication of uniform bulks with high reproducibility. Three homogeneous DyBa$_2$Cu$_3$O$_y$ ring-bulks with various sizes were synthesized in this study. All of these bulks exhibited concentrically cone-shaped trapped field distribution on the surface and high trapped field more than 1.8 T at 77 K inside the ring, the highest among bulks of comparable size to date. Furthermore, superconducting properties such as superconducting transitions and critical current densities are highly uniform throughout the bulk, confirming the effectiveness of the SDMG approach. Our findings represent a significant advancement in the direct fabrication of high-quality ring-shaped REBCO bulks suitable for high magnetic field applications, in particular, compact nuclear magnetic resonance systems.

REBa$_2$Cu$_3$O$_y$ (REBCO, RE: rare earth element) is a high temperature superconductor showing high critical temperature ($T_c$) of ~90 K, which is highter than the boiling point of liquid nitrogen, 77 K. Therefore, REBCO textured materials have been intensively developed owing to their high critical current density ($J_c$), particularly under high magnetic fields. There are two types of nuclear magnetic resonance (NMR) systems being developed utilizing REBCO materials. The first aims to create a flagship model using high-temperature superconducting tapes, including REBCO coated conductors, capable of generating a uniform and ultra-high magnetic field of ~30 T[1]. The second involves the development of compact, portable, and cryogen-free NMR systems using stacked ring-shaped REBCO melt-textured bulks, which can trap magnetic field as high as ~17 T[2–4]. In principle, these bulk NMR systems can be achieved with high performance levels equivalent to those of conventional large and stationary 200-400 MHz, *i.e.* 5–10 T, NMR systems. This approach offers substantial advantages, including significant miniaturization, improved portability, and the elimination of the need for liquid helium, making it an attractive and promising alternative for various applications. However, to utilize REBCO bulks for NMR applications, two critical requirements must be resolved: the growth of highly homogeneous REBCO bulks with high reproducibility and the realization of large-scale ring-shaped bulks without mechanical damages. The conventional crystal growth methods, such as the top-seeded melt growth (TSMG) and the top-seeded infiltration growth (TSIG), have made it challenging to grow homogeneous ring-shaped bulks, as the crystal growth initiates from a small seed crystal placed on top of the REBCO pellet. To create ring-shaped bulks suitable for NMR, mechanical processing on disk-shaped bulks to make a hole along the central axis is necessary as schematically shown in Fig. 1 (a). This process involves a risk of crack generation as well as results in significant material loss. A few instances of direct growth of ring-shaped bulks using the top-seed method have been reported, such as employing a multi-seed approach or placing a buffer pellet on a ring-shaped pellet[5–8]. Nevertheless, achieving a concentric trapped field in the ring-shaped REBCO bulk using the top-seeded method has been a significant challenge. Despite these hardles, T. Nakamura *et al.* have developed the world's first tabletop 200 MHz NMR systems using stacked EuBCO ring-shaped bulks fabricated by the top-seeded method. NMR signal linewidths of less than 1 ppm were obtained using this tabletop NMR system, demostrating its potential for portable NMR applications[9]. Given the promising potential of ring bulks for applications such as magnetic resonance imaging (MRI), high gradient trapped field magnets (HG-TFM), and magnetic shielding, in addition to desktop NMR, there is an increasing amount of research aiming to better understand their theoretical magnetization processes and mechanical characteristics[10–20].

In recent years, we have developed and reported on a novel bulk growth technique, the Single-Direction Melt Growth (SDMG) method, to address the challenges of scalability and homogeneity in the conventional top-seeded method[21,22]. In this approach, large bulk plates cut from commercially processed TSMG REBCO bulks, which exhibit relatively high peritectic temperature ($T_p$) such as EuBCO, are utilized as seed plates. REBCO bulks with a $T_p$ lower than that of the seed plate, such as YBCO, DyBCO, and GdBCO, are grown vertically from the seed plate. This strategy simplifies the conventional three-dimensional crystal growth into one-dimensional growth process confined to the vertical direction, making it easier to fabricate bulks with larger sizes and enabling the direct growth of complex shapes including ring-shaped structures. Figure 1(a) presents a schematic illustration of ring bulk growth using the SDMG method compared to the typical TSMG method. The SDMG method enables the direct growth of ring-bulks and they are separated simply by cutting from the seed plate, which can be reused multiple times. We are confident that the SDMG method is well-suited for the prepartion of REBCO ring-bulks for NMR applications. In the previous study, we successfully fabricated homogeneous REBCO bulks (RE of Y, Dy, and Gd) using the SDMG method, which demonstrated superior concentrically cone-shaped trapped field ($B_T$) distributions compared to those of TSMG processed bulks[22]. In the present study, we attempt the direct growth of ring-shaped bulks with varying diameters and thicknesses using the SDMG method and report on their $B_T$ characteristics and superconducting properties.

We selected DyBCO for the ring-shaped bulks and EuBCO as seed plates, respectively. The EuBCO seed plates with (001) plane on the surface were extracted from TSMG bulks (Eu-QMG®) with a diameter of ~64 mm provided by *Nippon Steel Co*. The precursor powders consisting of DyBCO, $Dy_2BaCuO_5$ (Dy211) with DyBCO : Dy211 = 7 : 3 in molar ratio, 10 wt% $Ag_2O$, and 0.5 wt% $CeO_2$ were provided from *TEP Co*. The $Ag_2O$ and $CeO_2$ were incorporated into the precursor to lower the $T_p$ and suppress the grain growth of Dy211, respectively. The precursor powder was then pressed into ring shapes using metallic dies under a uniaxial pressure of ~100 MPa. Subsequently, 'mm$^{OD}$', 'mm$^{ID}$', and 'mm$^t$' will refer to the outer diameter, inner diameter, and thickness, respectively. We utilized two types of metallic dies with (mm$^{OD}$, mm$^{ID}$) of (25 mm, 10 mm) and (35 mm, 10 mm) in this study. The precursor DyBCO rings underwent heat treatment above the peritectic temperature in air for a short time to ensure densification, followed by surface polishing. A viscous paste, created by combining the precursor powder and several organic constituents, was coated at the interface between the bulk and the seed plate to ensure uniform contact. All ring-shaped DyBCO bulks in this study were directly grown using the SDMG method in air, where the highest holding temperature must be between the $T_p$s of DyBCO (~990°C) and EuBCO (~1030°C) considering that the addition of Ag lowers the $T_p$s by approximaterly 20°C. For detailed SDMG conditions concerning Ag-added DyBCO, please refer to our previous study[22]. In total, three Ag-added DyBCO bulks (numbered #Dy1–3, hereinafter) with 25 and 35 mm$^{OD}$, were fabricated in this study. Appearance of the prepared bulks is shown in Fig.1 (b) and their characteristics, such as the size and trapped fields, are summarized in Table I. After separating the grown bulks from the seed plate using a diamond saw, post-annealing under the moderately reductive condition, *i.e.* 1% $O_2$/Ar flow at 900°C for 24 h, was carried out for the bulks, followed by oxygen annealing under flowing oxygen gas at 425°C for more than 200 h to control carrier doming state being optimal. Post-annealing conditions were decided based on our recent study on the effects of reductive annealing on REBCO melt-grown bulks[23]. Using the SDMG method, approximately 1 mm of randomly oriented regions remains in the uppermost part, *i.e.* growth front, of the bulks due to the accumulation of extruded RE211 grains. Therefore, 1–2 mm from the growth front was removed for all prepared bulks. Notably, all bulks fabricated using the SDMG method exhibit a slightly truncated-cone shape. Despite the diameters listed in Table I referring to the seed side's outer and inner dimensions, the diameters on the growth front side are 1-3% smaller. $B_T$ distributions ~1 mm above the bulk-surfaces and $B_T$ changes in the depth direction within the center of the ring were evaluated at 77 K using a Hall probe (*Lake Shore Cryotronics Inc.* HGCA-3020) after field cooling magnetization up to 2 T. The $B_{T, \max}$ represents the highest $B_T$ value inside the ring. Zero-field cooled magnetization under 10 Oe and magnetization hysteresis loops at 77 and 60 K under dc fields up to 5 T were measured for small rectangular pieces cut from various positions of the #Dy2 bulk using a SQUID magnetometer (Quantum Design MPMS XL-5s). Magnetic fields were always applied parallel to the *c*-axis of the samples. $J_c$ values were calculated from the widths of magnetization hysteresis based on the extended Bean model.

For all samples, #Dy1–#Dy3, we successfully grew crack-free ring-shaped REBCO bulks as shown in Fig. 1, which exhibited the glossy appearance characteristic of melt-textured bulks. Figure 2 (a) shows the $B_T$ distribution at 77 K above seed side and growth front surfaces of the #Dy1 bulk. Both surfaces demonstrate high $B_T$ distributions in the shape of concentrical cone with almost equivalent values, indicating that a uniform bulk material has been obtained in both the radial and height directions. Figures 2 (b) and (c) depict the two-dimensional $B_T$ distribution above the seed side surface of the #Dy3 bulk and the estimated in-plane current density (*J*) distribution, respectively. A homogeneous cone-shaped distribution, along with high $B_T$ value of more than 0.9 T, is also observed. *J* distributions were calculated by resolving the inverse Biot-Savart problem using the previously proposed inverse matrix method[24,25]. In this calculation, we only adopted two assumptions: currents flowing only in the *ab* plane and a constant *J* along the thickness direction. Without any boundary conditions, we obtained a circulation of *J* reflecting the ring shape of the bulk. In addition, to quantitatively evaluate the homogeneity of the prepared bulks, we computed the circularity and ellipticity for the $B_T$ distribution above the seed side, as shown in Table I. Circularity (*c*) and ellipticity (*e*) were calculated for isomagnetic field lines at 0.02 T intervals using the following equations: $c = 4\pi S/L^2$ and $e = 1 - b/a$, where *L*, *S*, *a*, and *b* represent the perimeter, area, major axis, and minor axis of each closed isomagnetic field line, respectively. The mean values were then calculated. The closer *c* is to 1 and *e* is to 0, the closer the obtained $B_T$ distribution approximates a perfect circle. A more detailed explanation of these parameters can be found in reference[22]. All ring-shaped bulks exhibited *c* and *e* values remarkably close to a perfect circle, with *c* > 0.98 and *e* < 0.05, whose values are equivalent to, or even surpass those of homogeneous disk-shaped SDMG bulks reported thus far[22]. For desktop NMR applications, where a larger sample space is desired, further scaling-up is expected. The SDMG method is advantageous for enlargement since the crystallization process progresses only in the vertical direction and the crystallization time is independent of the size in the radial direction in principle.

In fact, we have successfully grown homogeneous ring-shaped bulks up to 50 mm$^{OD}$ × 25 mm$^{ID}$, details of which will be provided in an upcoming publication.

Figure 3 shows the $B_T$ distribution at 77 K in the height direction along the center of rings for #Dy1–3. These samples demonstrated extremely high trapped magnetic fields exceeding 1 T, with #Dy3, in particular, exhibiting a central magnetic field surpassing 1.8 T. To the best of our understanding, the $B_T$ value at 77K in our case is the highest among bulks of a comparable size. O. Vakaliuk *et al.* reported a $B_T$ value of 1.13 T at 77K inside a YBCO ring bulk similar in size to #Dy3, *i.e.* 28.1 mm$^{OD}$ × 6.0 mm$^{ID}$ × 17.3 mm$^{t}$, and this ring-shaped bulk trapped up to 9.78 T when cooled to 25 K[26]. We anticipate that cooling our bulk to low temperatures will enable the realization of an internal magnetic field exceeding 10 T.

Finally, superconducting properties were evaluated in detail for the prepared ring-bulk, #Dy2. As schematically shown in Fig. 4 (a), small pieces were cut from different positions, namely varying radial positions (inner side, center, and outer side) and different heights, within the #Dy2 sample, and their superconducting properties were evaluated. Figures 4 (b) and (c) summarize the sample position dependences of transition widths ($\Delta T_c$), $T_c^{mid}$, and $J_c$ at 77 K, 1 T and 60 K, 3 T, respectively. $\Delta T_c$ is defined as the temperature difference between $T(\chi = -0.05)$ and $T(\chi = -0.95)$, while $T_c^{mid}$ is defined as the temperature at which $\chi = -0.5$, where $\chi$ is zero-field-cooled magnetization normalized at 80 K. Despite slight variation in $T_c^{mid}$ between 90.8–91.9 K in the radial direction, all samples exhibited very sharp superconducting transitions with the transition widths less than 1.5 K, even at the positions distant from the seed. This suggests that the RE/Ba substitution, which deteriorates both $T_c$ and $\Delta T_c$, is suppressed and the carrier doping state is almost uniform throughout the bulk. In Fig. 4 (c), the distribution of $J_c$ measured from various positions in the commercial EuBCO bulk is also indicated by bands. Although this sample is not identical to the seed plate used in this study, it was produced using the same growth method and is a homogeneous product also used in the compact bulk NMR that Nakamura *et al.* has reported[9]. From the $J_c$ characteristics of the SDMG bulks, the overall performance in magnetic fields is the same or higher with an equivalent level of uniformity as the commercial bulk. It should be noted that there is a strong correlation between $J_c$ at 77 K, 1 T and that at 60 K, 3 T, suggesting that the $J_c$ characteristics at middle to low temperatures, where desktop NMR applications are expected to operate, can be inferred from the in-field $J_c$ properties at 77 K.

In conclusion, we have succeeded in the direct fabrication of high-quality ring-shaped REBCO melt textured bulks using the single-direction melt growth (SDMG) method, which overcomes the shape limitations posed by conventional top-seeded crystal growth methods. The prepared bulks demonstrated concentrically cone-shaped trapped field distributions on the surface and impressively high trapped fields exceeding 1.8 T at 77 K inside the center of the ring, which is the highest record among ring-shaped bulks of comparable size to date. Moreover, these bulks exhibited highly uniform superconducting properties across both radial and height dimensions. These results strongly highlight the promising potential of REBCO ring-shaped bulks fabricated using the SDMG method, which opens up innovative applications, particularly in the development of compact, cryogen-free NMR systems. Continued enhancement of the SDMG method, combined with scaling up of bulk sizes will facilitate the realization of these promising applications.

This research was supported by NEDO Uncharted Territory Challenge 2050 (22M1C01Y) and by research grant program of Nippon Sheet Glass Foundation for Materials Science and Engineering.

**Figures and Tables**

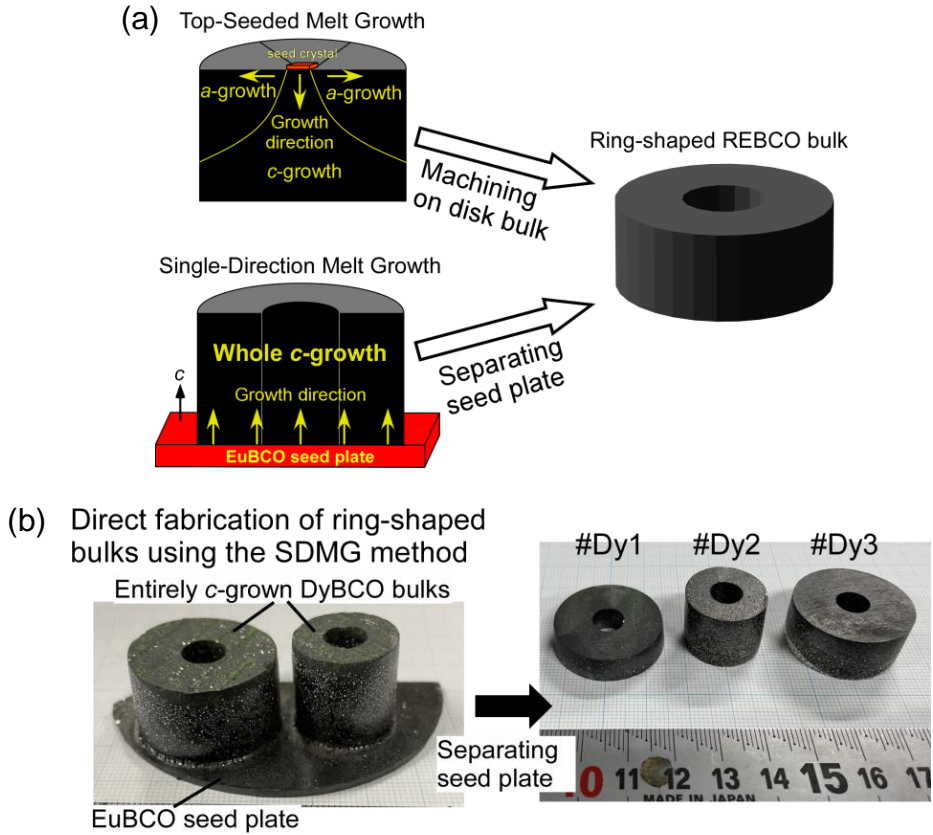

FIG. 1. Schematic illustration comparing the fabrication processes of a ring-shaped REBCO bulk via the top-seeded melt growth (TSMG) or the single-direction melt growth (SDMG) method (a) and appearance of the directly prepared DyBCO ring-bulks, #Dy1–3, using the SDMG method in this study (b), corresponding to the illustration of SDMG process in (a).

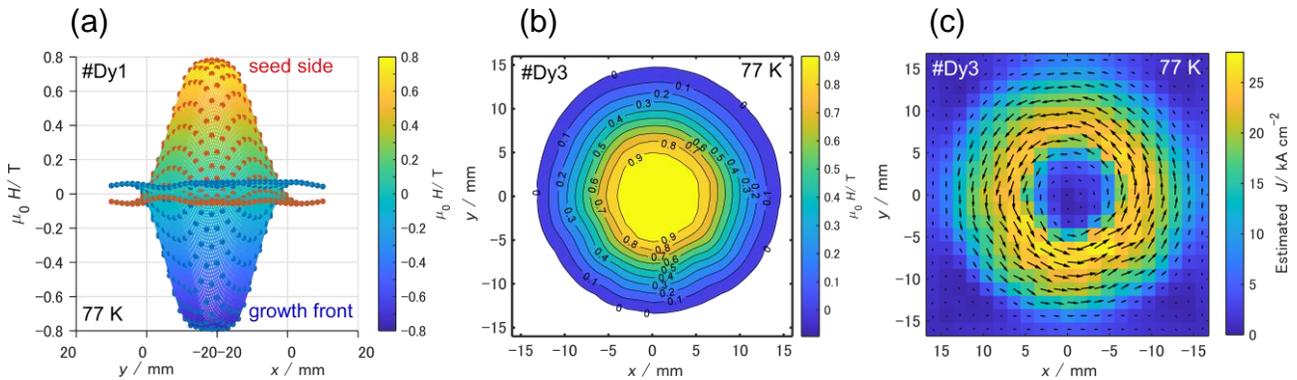

FIG. 2. Three-dimensional trapped magnetic field distributions at 77 K above the seed side and growth front surfaces of the #Dy1 bulk (a), where both surfaces exhibit concentrically cone-shaped trapped magnetic fields, indicating a uniform bulk material in radial and height directions. Two-dimensional trapped magnetic field distribution at 77 K above the seed side surface of the #Dy3 bulk (b) and computed in-plane current density distribution obtained by solving the inverse Biot-Savart problem (c).

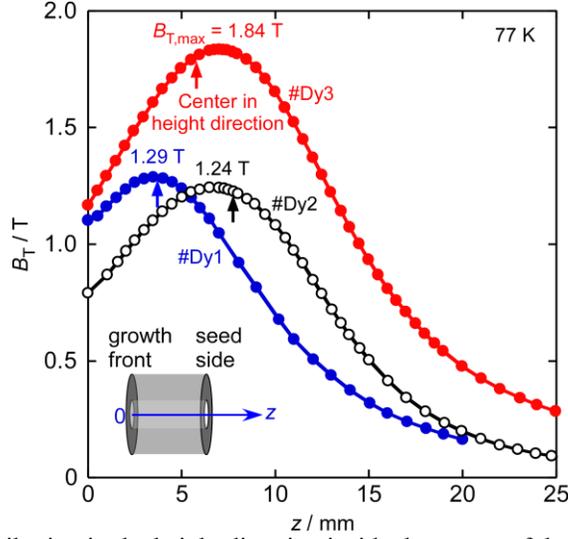

FIG. 3. Trapped magnetic field distribution in the height direction inside the center of the rings of the #Dy1-#Dy3 bulks. Samples exhibit extremely high trapped magnetic fields exceeding 1 T, with #Dy3 in particular showing a central magnetic field surpassing 1.8 T.

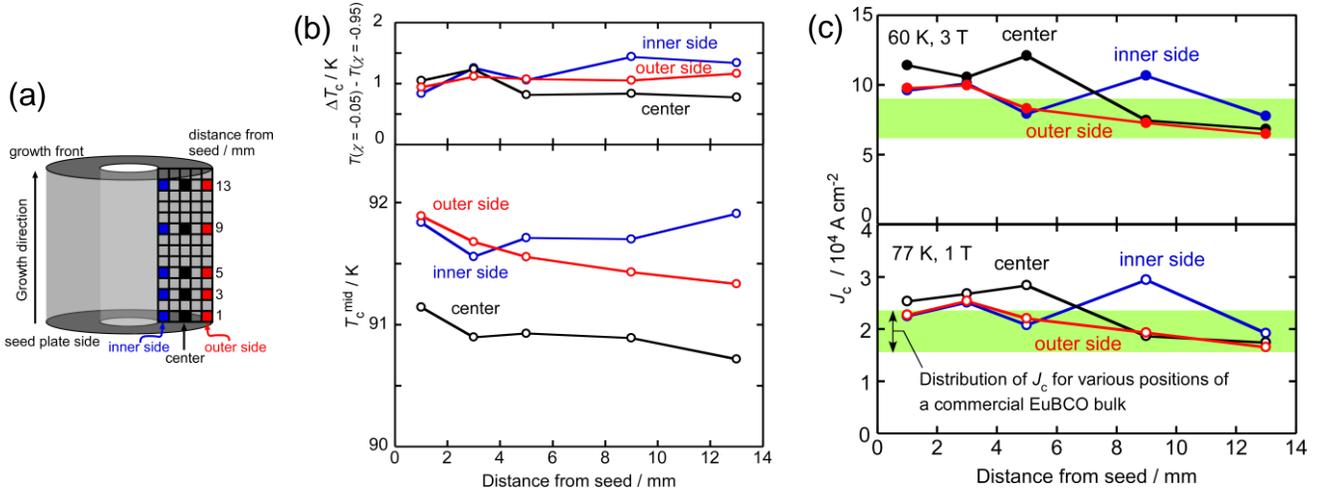

FIG. 4. Schematic illustration of the measured sample positions (a), sample position dependences of $\Delta T_c$ and $T_c^{mid}$ (b), and of $J_c$ at 77 K, 1 T (open) and 60 K, 3 T (closed) (c). $\Delta T_c$ is defined as temperature difference between $T(\chi = -0.05)$ and $T(\chi = -0.95)$, while $T_c^{mid}$ is defined as the temperature at which $\chi = -0.5$, where $\chi$ is a zero-field cooled magnetization under 10 Oe normalized at 80 K. Distribution of $J_c$ measured at various positions in the commercial EuBCO bulk is also indicated in the figure (c) by green bands.

TABLE I. Summary of the size of die and grown bulks, $B_{T,max}$ at 77 K, circularity, and ellipticity for the prepared ring-shaped DyBCO bulks in this study. *OD* and *ID* represent the outer diameter and inner diameter, respectively.

| No. | Die-size<br>*OD* × *ID* / mm² | Grown bulk-size<br>*OD* × *ID* × thickness / mm³ | $B_{T,max}$ (77 K) / T | Circularity, *c* | Ellipticity, *e* |
|---|---|---|---|---|---|
| #Dy1 | 35 × 10 | 29.0 × 8.3 × 7.4$^t$ | 1.29 | 0.989 | 0.022 |
| #Dy2 | 25 × 10 | 20.1 × 7.5 × 15.4$^t$ | 1.24 | 0.991 | 0.031 |
| #Dy3 | 35 × 10 | 28.7 × 7.9 × 11.5$^t$ | 1.84 | 0.987 | 0.040 |